\documentclass[onecolumn,showpacs]{revtex4}

\usepackage{graphicx}
\usepackage{dcolumn}
\usepackage{bm}

\input epsf

\begin{document}

\title{\Large Black Hole Thermodynamics in Ho$\breve{r}$ava Lifshitz Gravity and the Related Geometry }

\author{\bf Ritabrata
Biswas\footnote{biswas.ritabrata@gmail.com}~And~Subenoy
Chakraborty.\footnote{schakraborty@math.jdvu.ac.in} }

\affiliation{Department Of Mathematics, Jadavpur University}

\date{\today}

\begin{abstract}
Recently, Ho$\breve{r}$ava proposed a non-relativistic
renormalizable theory of gravity which is essentially a field
theoretic model for a UV complete theory of gravity and reduces to
Einstein gravity with a non-vanishing cosmological constant in IR.
Also the theory admits a Lifshitz scale-invariance in time and
space with broken Lorentz symmetry at short scale. On the other
hand, at large distances higher derivative terms do not contribute
and the theory coincides with general relativity. Subsequently,
Cai and his collaborators and then Catiuo et al have obtained
black hole solutions in this gravity theory and studied the
thermodynamic properties of the black hole solution. In the
present paper, we have investigated the black hole thermodynamic
for two choices of the entropy function - a classical and a
topological in nature. Finally, it is examined
whether a phase transition is possible or not.\\

\end{abstract}


\maketitle

\section{INTRODUCTION}

Recently, a new four dimensional gravity theory has been proposed
by Horava (2009) without full diffeomorphism invariance but UV
completeness. In fact, the theory has three dimensional general
covariance and time re-parameterization invariance. Essentially,
it is a non-relativistic renormalizable quantum gravity theory
with higher spatial derivatives. Due to anisotropic rescaling
Horava's theory is power-counting renormalizable. At large
distance the general covariance is recovered and the theory
reduces (with some restrictions) to Einstein gravity with a
non-vanishing cosmological constant in IR , i.e., the general IR
vacuum of this theory is anti-deSitter. However, if one adds a
new term ($\mu^{4}R$)in the action and take $\Lambda\rightarrow 0$
then although UV properties of the theory do not change but IR
properties alters and one gets Minkowskin vacuum in the IR
region. Moreover, due to Lifshitz scale invariance $(t\rightarrow
l^{z}t, x^{i}\rightarrow l x^{i}, z\geq 1 )$ of the space time
this theory is also known as Horava-Lifshitz theory (HL). But in
semiclassical treatment of scalar excitation of HL gravity shows
the usual naturalness problem of Lorentz violating theories.

Horava has used ADM formalism where the four dimensional metric is
parameterized as (Arnowitt et. al. 1962)
\begin{equation}
ds^{2}=-N^{2}dt^{2}+g_{i
j}\left(dx^{i}+N^{i}dt\right)\left(dx^{j}+N^{j}dt\right)
\end{equation}
Here the lapse function $N$, shift vector $N^{i}$ and the
3-Dimensional spatial metric $g_{i j}$ are the dynamical
variables in Horava-Lifshitz gravity. So the Eintein-Hilbert
action has the form
\begin{equation}
S=\frac{1}{16 \pi G}\int d^{4}x \sqrt{g}~ N\left\{ \left(K_{i
j}K^{i j}-K^{2}\right)+R-2 \Lambda\right\}
\end{equation}
where$$G=~Newton's~gravitational~constant,$$ $R$ is the curvature
scalar for the 3-metric $g_{i j}$ and the extrinsic curvature
$K_{i j}$ is defined as
\begin{equation}
K_{ij}=\frac{1}{2 N}\left(\dot{g}_{i
j}-\nabla_{i}N_{j}\nabla_{j}N_{i}\right)
\end{equation}
 Here an over dot stands for time derivative while $\nabla_{i}$
 is the covariant derivative with respect to the spatial three
 metric $g_{i j}$.

The action of the non-relativistic renormalizable theory proposed
by Ho$\breve{r}$ava (2009) (known as Ho$\breve{r}$ava-Lifshitz (HL)
action) has the expression
\begin{equation}
I=\int dt~ d^{3}x \sqrt{g}~ N \left(L_{0}+~L_{1}\right)
\end{equation}
with
$$L_{0}=\left\{\frac{2}{\kappa^{2}}\left(K_{i j}K^{i j}-\lambda K^{2}\right)+\frac{\kappa^{2}\mu^{2}
\left(\Lambda R-3
\Lambda^{2}\right)}{8\left(1-3\lambda\right)}\right\}$$
$$L_{1}=\left\{\frac{\kappa^{2}\mu^{2}
\left((1-4\lambda\right)}{32\left(1-3\lambda\right)}R^{2}-\frac{\kappa^{2}}{2\omega^{4}}Z_{i
j}Z^{i j}\right\}$$

Here
\begin{equation}
Z_{i j}=C_{i j}-\frac{\mu ~\omega^{2}}{2}R_{i j}~~,
\end{equation}

the Cotten tensor $C^{i j}$ has the expression
\begin{equation}
C^{i j}=\epsilon^{i k l}\nabla_{k}\left(R_{l}^{j}-\frac{1}{4}R ~
\delta_{l}^{j}\right) =\epsilon^{i k
l}\nabla_{k}R_{l}^{j}-\frac{1}{4}\epsilon^{i k j}\partial_{k}R
\end{equation}
and $\kappa^{2},~ \lambda,~ \mu,~ \omega$ and $\Lambda$ are
constant parameters.

In the above Ho$\breve{r}$ava-Lifshitz action $(4)$ the first two
terms are the kinetic terms and the rest correspond to potential
of the theory (in $'detailed- balance'$ form ). Now comparing with
general relativity action, the expressions for the speed of light,
Newton's constant and the cosmological constant are
\begin{equation}
c=\frac{\kappa^{2} \mu}{4}\sqrt{\frac{\Lambda}{1-3
\lambda}}~,~G=\frac{\kappa^{2}c}{32
\pi}~,~\tilde{\Lambda}=\frac{3}{2}\Lambda~.
\end{equation}
Note that in the present theory (Horava 2009) $\lambda$ is a dynamical
coupling constant, subject to quantum correction. In fact, for
$\lambda=1$ the first three terms in the action $(4)$ can be
converted to the usual ones in Einstein's general relativity. Also
the expression for the velocity of light (in equation $(7)$)
demands the cosmological constant $\Lambda$ must be negative if
$\lambda> \frac{1}{3}$. However, an analytic continuation (Lu et. al. 2009)
\begin{equation}
\mu\rightarrow i \mu,~~~~ \omega^{2}\rightarrow -i \omega^{2}
\end{equation}
keeps the action $(4)$ to be real (Calcagni 2009) and we may choose
$\Lambda$ to be positive for $\lambda>\frac{1}{3}$. The
cosmological implications of the HL action has been studied by
Kiritsis et. al. (2009), Ho$\breve{r}$ava (2009; 2009), Takahasi et. al. (2009), Kluson (2009), Cai et. al. (2009; 2009).

Now variation of the HL action with respect to N, $N^{i}$ and
$g_{i j}$ give the equations of motion
\begin{equation}
\frac{2}{\kappa^{2}}\left(K_{i j}K^{i j}- \lambda K^{2}\right)
-\frac{\kappa^{2} \mu^{2} \left(\Lambda R-3
\Lambda^{2}\right)}{8\left(1-3 \lambda\right)} -\frac{\kappa^{2}
\mu^{2} \left(1-4 \Lambda\right)}{32\left(1-3
\lambda\right)}R^{2}+\frac{\kappa^{2}}{2 \omega^{4}}Z_{i j}Z^{i
j}=0
\end{equation}
\begin{equation}
\nabla_{k}\left(K^{k l}-\lambda K g^{k l}\right)=0
\end{equation}
$$and$$
\begin{equation}
\frac{2}{\kappa^{2}}E_{i j}^{(1)}- \frac{2
\lambda}{\kappa^{2}}E_{i j}^{(2)}+\frac{\kappa^{2} \mu^{2}
\Lambda}{8\left(1-3 \lambda\right)}E_{i j}^{(3)}+\frac{\kappa^{2}
\mu^{2} \left(1-4 \lambda\right)}{32\left(1-3 \lambda\right)}E_{i
j}^{(4)}-\frac{\mu \kappa^{2}}{4 \omega^{2}}E_{i j}^{(5)}- \frac{
\kappa^{2}}{2 \omega^{4}}E_{i j}^{(6)}=0
\end{equation}
where the tensors $E_{i j}^{(\alpha)}~~~(\alpha = 1, 2,.....,6)$
are the combination of $K_{i j},~g_{i j},~N,~N_{i}$ and their
covariant derivative with respect to the three dimensional metric
and detailed expressions can be found by Arnowitt et. al. (1962).

Immediately, after the proposal of this new theory, Lu et. al. (2009)
have obtained spherically symmetric solutions in HL gravity and
have shown asymptotically $AdS_{2}$ solution for $\lambda=1$.
Subsequently, their solution has been extended to general
topological black hole solution by Cai et al (2009) and they have
studied extensively the corresponding black hole thermodynamics.
Subsequently the HL theory has been modified by introducing a new
form $\mu^{4}R$ in the action. Then Kehagias and Sfetsos (2009) has
obtained asymptotic flat spherically symmetric vacuum black hole
solution (known as KS black hole). For spherically symmetric
solution considering $N_{i}=0$ in $(1)$ we obtain the metric
ansatz
\begin{equation}
ds^{2}=-N^{2}(r)dt^{2}+\frac{dr^{2}}{f(r)}+r^{2}d^{2}\Omega
\end{equation}
From this line element evaluating the angular integration, the
modified Lifshitz-Horava Lagrangian reduces to
\begin{equation}
\tilde{{\cal
L}}_{1}=\frac{\kappa^{2}\mu^{2}N}{8\left(1-3\lambda\right)\sqrt{f}}
\left\{\frac{\lambda-1}{2}f'^{2}+\frac{\left(2\lambda-1\right)
\left(f-1\right)^{2}}{r^{2}}-\frac{2\lambda\left(f-1\right)}{r}f'-2\omega\left(1-f-rf'\right)\right\}
\end{equation}
with
$$\omega=\frac{8\mu^{2}\left(3\lambda-1\right)}{\kappa^{2}}$$
Now $\lambda=1$ gives $\omega=\frac{16\mu^{2}}{\kappa^{2}}$ and
then the field equations obtained from the Lagrangian $(13)$, can
be solved to obtain $'f'$ and $'N'$ as (Kehagias et. al. 2009; Catillo et. al. 2009) the values of $f$
and $N$ can be evaluated as (Kehagias et. al. 2009)
\begin{equation}
N^{2}=f(r)=1+\omega r^{2}-\sqrt{r\left(\omega^{2}r^{3}+4\omega
M\right)}
\end{equation}
In the above solution the integration constant $'M'$ is related to
the mass of the corresponding black hole. In fact, if $r_{+}$be
the radius  of the event  horizon then
\begin{equation}
M=\frac{r_{+}}{2}+\frac{1}{4\omega r_{+}}
\end{equation}

This expression for $M$ is similar to the mass-charge relation for
(non-extremal) Reissner-Nordstrom black hole. Here
$\frac{1}{\sqrt\omega}$ corresponds to charge parameter in
Reissner-Nordstrom solution and is taken as a new parameter in the black hole thermodynamics. If
$M^{2}>\frac{1}{2\omega}$ then there are more than one horizon while there will be degenerate
horizon at $r_{c}=M=\sqrt{\frac{1}{2\omega}}$(extremal one) and no horizon exists for $M^{2}<\frac{1}{2\omega}$
and we are left with a naked singularity.

The nice similarity between black hole and thermodynamical system
was first introduced by Bekenstein (1973) and Hawking (1975), by
relating the surface gravity of the horizon of the black hole to
the temperature of the thermodynamical system(known as Hawking
temperature) while the area of the horizon is proportional to the
entropy of the thermodynamical system,i.e.,
\begin{equation}
T=\frac{\kappa}{2\pi}~~,~~S=\frac{A}{4}
\end{equation}
and these entropy and temperature are related by the first law of
thermodynamics (Bardeen et. al. 1973). But in general the area formula for entropy
breaks down for higher derivative gravity theories. Using first
law of thermodynamics we can write entropy as
$$S=\int \frac{d M}{T}+S_{0}.$$

The integration constant $'S_{0}'$ should be fixed by physical
consideration. As the mass of a black hole is a function of the
radius of the horizon $r_{+}$ so we can write the above integral
as

$$S=\int \frac{1}{T}\frac{\partial M}{\partial r_{+}}dr_{+}+S_{0}$$

Using $M$ from $(15)$ we have
\begin{equation}
S=\pi r_{+}^{2}+\frac{2\pi}{\omega}\ln r_{+}
\end{equation}

(choosing the integration constant $S_{0}=0$ to match with
Schwarzschild black hole) This is expression of entropy of the HL
black hole using Hawking temperature and first law of
thermodynamics on the horizon. Also it is the usual Bekenstein
entropy with logarithmic correction. This entropy is similar to
that for topological black holes obtained by Cai et. al.(2009). In
this connection, one may note that in GR logarithmic corrections
in entropy is due to thermodynamic fluctuations around thermal
equilibrium.Also the logarithmic correction term is expected to be
a generic one in any theory of quantum gravity.

 Usually, a black hole is characterized by
three parameters namely its mass, charge and angular
momentum(known as no hair theorem) and the thermodynamical
stability of the black hole is determined by the sign of its heat
capacity$(c_{\omega})$. If $c_{\omega}<0$ (as Schwarzschild black
hole), then black hole is thermodynamically unstable, but if
$c_{\omega}$ changes sign in the parameter space such that it
diverges (Hut 1977) in between then from ordinary thermodynamics it
indicates a second order phase transition (Davies 1977; 1977; 1989). However in black
hole thermodynamics, a critical point exits also for extremal
black hole and the second order phase transition takes place from
an extremal black hole to its non extremal counter part.

    On the other hand the geometrical concept into ordinary thermodynamics was first
introduced by Weinhold (1975) (for details see the review (Ruppeiner 1995; 1996)).
According
 to him a Riemannian metric can be defined as the second derivative of
the internal energy ($U$) to entropy ($S$) and other extensive
variables ($y^{\alpha}$) of the system ,i.e., the Hessian of the
energy is known as Weinhold metric
\begin{equation}
g_{i j}^{(W)} = \partial_{i}\partial_{j}U(S,y^{\alpha})
\end{equation}
However it looses physical meaning in equilibrium thermodynamics.
Subsequently, Ruppeiner (1979) introduced a metric (known as the
Ruppeiner metric) as the Hessian matrix of the thermodynamic
entropy. It is defined on the state space as
\begin{equation}
g_{i j}^{(R)} = -\partial_{i}\partial_{j}S(U,y^{\alpha})
\end{equation}
The Ruppeiner metric and the Weinhold metric are conformally
related as
\begin{equation}
ds_{R}^{2} = \frac{1}{T}ds_{W}^{2}
\end{equation}
Unlike Weinhold metric, the Ruppeiner geometry has physical
relevance in the fluctuation theory of equilibrium thermodynamics.

In the present work, we make a comparative study of the black hole
thermodynamics for the above modified Horava-Lifshitz gravity with
logarithmically corrected entropy (called as entropy-$I$) given by
equation $(17)$ and the usual area-entropy relation (called as
entropy-$II$) given in equation $(16)$. The paper is organized as
follows : In section $II$ black hole thermodynamics has been
studied for entropy-$I$ and the variation of the thermodynamical
quantities has been examined. A similar analysis with entropy-$II$
has been presented in section $III$.The paper ends with a
conclusion in section-$IV$.

\section{Black hole Thermodynamics in Horava-Lifshitz Gravity with Entropy-I }

\subsection{Calculations}

If $r_{+}$ denotes the event horizon then from equation $(14)$ the
mass, entropy are given by equation $(15)$ and $(17)$ and other
thermodynamical quantities namely temperature, specific heat and
free energy are given by,

\begin{equation}
\bullet~~~T=\left(\frac{\partial M}{\partial
S}\right)_{\omega}=\frac{2r_{+}^{2}\omega-1}{8\pi
r_{+}\left(1+\omega r_{+}^{2}\right)}
\end{equation}

\begin{figure}
\includegraphics[height=3in, width=3in]{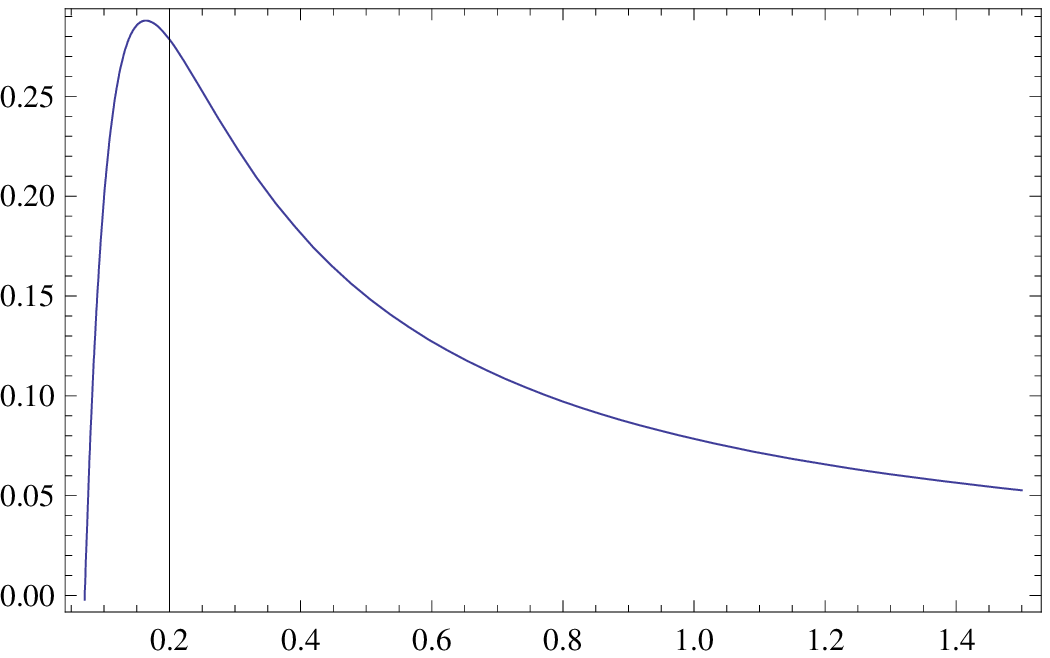}
\includegraphics[height=3in, width=3in]{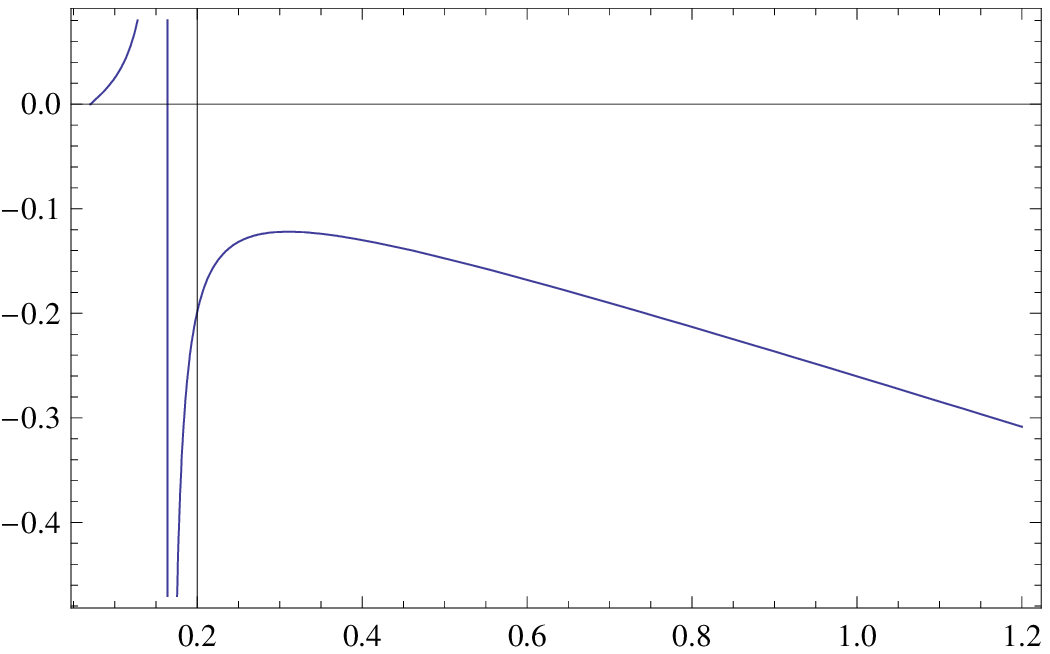}
Fig.(1a)~~$T_{H}$ vs $r_{+}$ graph, $\omega$ =
100.~~~~~~~~~~~~~~~~~~~~~~~~~~~~~~~~~~~Fig.(1b)~~$c_{\omega}$ vs
$r_{+}$
graph, $\omega$ =100~~~~~~~~~~~~~~~~~~~~~~~~~~\\

\vspace{5mm}
\includegraphics[height=3in, width=3in]{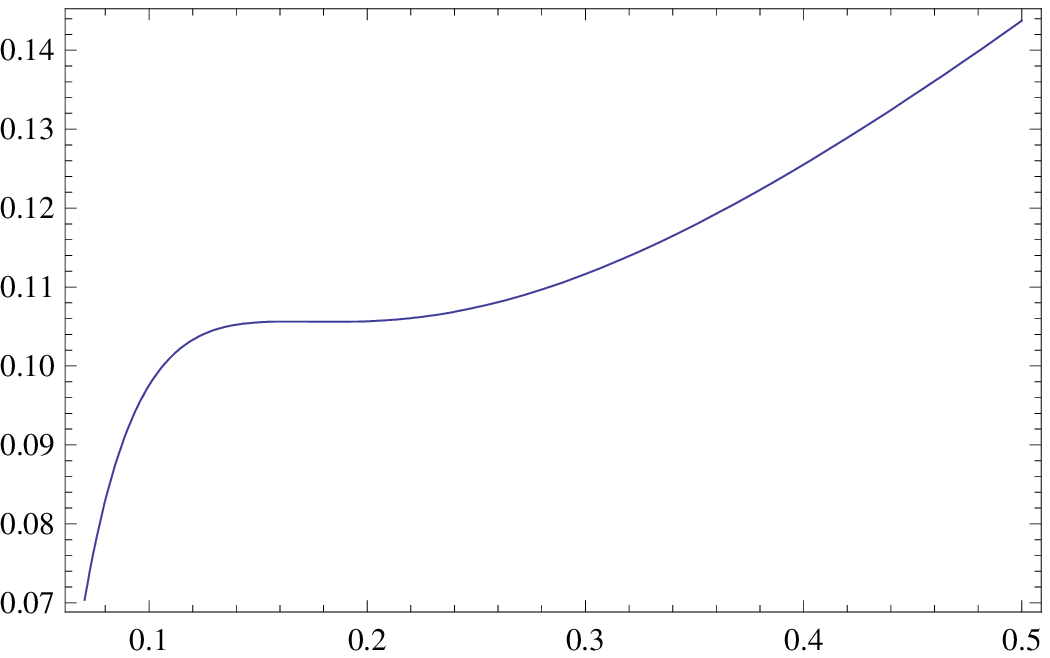}
\includegraphics[height=3in, width=3in]{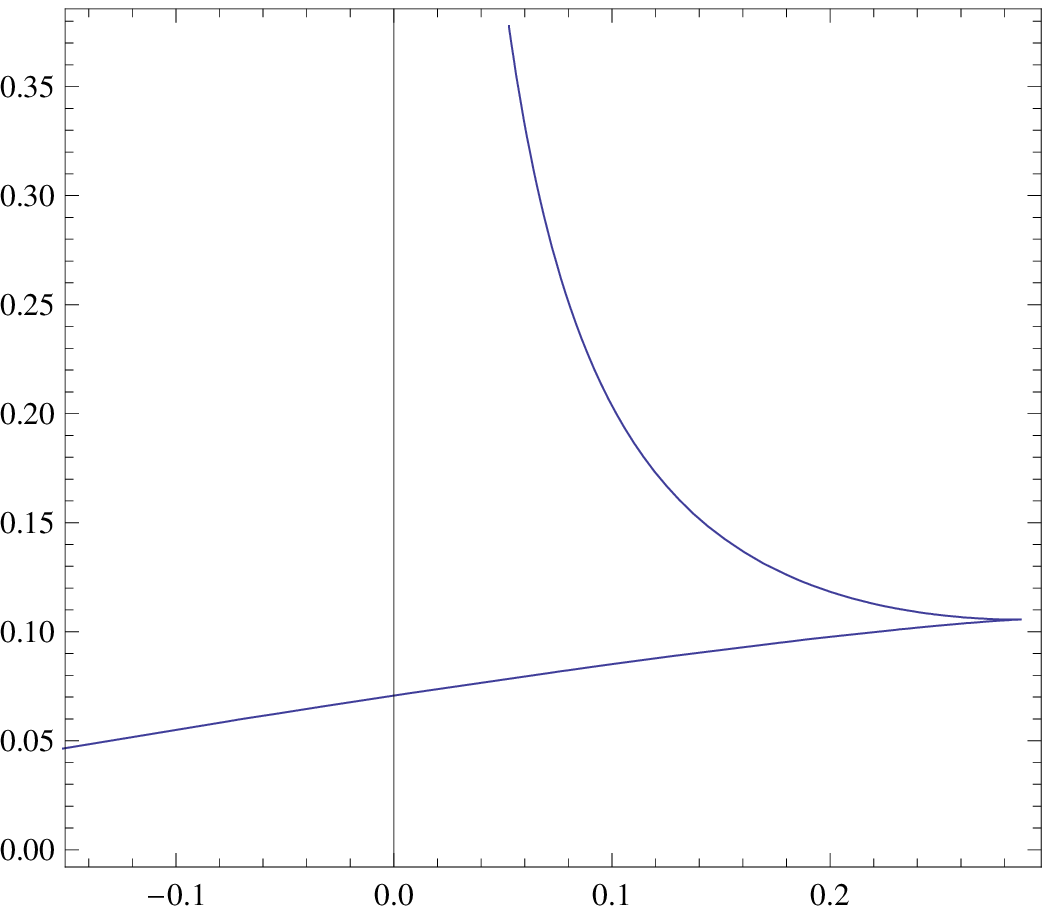}
Fig.(1c)~~$F$ vs $r_{+}$ graph, $\omega$ =
100.~~~~~~~~~~~~~~~~~~~~~~~~~~~~~~~~~~Fig.(1d)~~$F$ vs $T_{H}$ graph. $\omega$ =
100.~~~~~~~~~~~~~~~~~~~~~~~~\\
\hspace{1cm} \vspace{1mm} Fig. 1(a), 1(b), 1(c) show the variation
of $T_{H}$, $c_{\omega}$ and $F$ respectively with $r_{+}$. Fig.
1(d) represents the variation of $F$ with respect to $T_{H}$
\vspace{5mm}
\end{figure}

\begin{equation}
\bullet~~~c_{\omega}=T\left(\frac{\partial S}{\partial
T}\right)_{\omega}=\frac{2\pi}{\omega}\frac{\left(1+\omega
r_{+}^{2}\right)^{2}\left(2\omega
r_{+}^{2}-1\right)}{\left(1+5\omega r_{+}^{2}-2\omega^{2}
r_{+}^{4}\right)}
\end{equation}
\begin{equation}
\bullet~~~Free~~ energy ~~(F)
=M-S.T=\frac{2+r_{+}^{2}\omega\left(7+2r_{+}^{2}\omega\right)+2\left(1-2r_{+}^{2}\omega\right)\ln
r_{+}} {8r_{+}\omega\left(1+r_{+}^{2}\omega\right)}
\end{equation}
Now to derive the Weinhold metric we choose $'\omega'$ as the
other extensive variable of the thermodynamic system. The reason
for choosing $'\omega'$ as a parameter is that it corresponds to
charge in Reissner-Nordstrom solution (mentioned in the
introduction). Also recently, Wang et al (2009) have treated
$'\omega'$ as a new state parameter in studying the first law of
thermodynamics in IR modified HL space time. Thus Weinhold metric
has the form,

\begin{equation}
dS_{W}^{2}=\frac{\omega}{16\pi^{2}r_{+}}\frac{\left(1+5r_{+}^{2}\omega-2r_{+}^{4}\omega^{2}\right)}
{\left(1+r_{+}^{2}\omega\right)^{3}}~dS^{2}+\frac{3}{4\pi}\frac{r_{+}}{\left(1+r_{+}\omega\right)^{2}}~dSd\omega
+\frac{1}{2\omega^{3}r_{+}}~d\omega^{2}
\end{equation}
and the Ruppeiner metric is
\begin{equation}
dS_{R}^{2}=\frac{1}{T}dS_{W}^{2}=-\frac{\omega}{2\pi}\frac{\left(1+5r_{+}^{2}\omega-2r_{+}^{4}\omega^{2}\right)}
{\left(1-3r_{+}^{4}\omega^{2}-2r_{+}^{6}\omega^{3}\right)}~dS^{2}+
\frac{6r_{+}^{2}}
{\left(2r_{+}^{4}\omega^{2}+r_{+}^{2}\omega-1\right)}~dSd\omega+
\frac{4\pi}{\omega^{3}}\frac{\left(1+\omega
r_{+}^{2}\right)}{\left(2\omega r_{+}^{2}-1\right)}d\omega^{2}
\end{equation}
and after diagonalizing we get,
\begin{equation}
dS_{R}^{2}=-\frac{\omega}{4\pi}\frac{ \left(2+12\omega
r_{+}^{2}-3\omega^{2}r_{+}^{4}-4\omega^{3}r_{+}^{6}\right)}
{\left(1-2r_{+}^{2}\omega\right)\left(1+r_{+}^{2}\omega\right)^{3}}dS^{2}
+d\sigma^{2}
\end{equation}
where $\sigma$ can be determined from the equation
\begin{equation}
d\sigma=\sqrt{\frac{\pi}{\omega^{3}}\frac{\left(2\omega
r_{+}^{2}-1\right)}{\left(1+\omega
r_{+}^{2}\right)}}\left[\frac{\left\{2+\omega
r_{+}^{2}\left(4-3\ln
r_{+}\right)+2\omega^{2}r_{+}^{4}\right\}}{\left(2\omega^{2}r_{+}^{4}+\omega
r_{+}^{2}-1\right)}d\omega+\frac{3\omega^{2}r_{+}}{\left(2\omega
r_{+}^{2}-1\right)}dr_{+}\right]
\end{equation}

\subsection{Graphical Analysis and Physical Interpretations}

We shall now analyze the thermodynamical quantities graphically.
The Hawking temperature ($T$) is zero at $r_{+}=r_{+1}$(say) and
then increases sharply to a maximum $T_{M}$ at $r_{+}=r_{+2}$
(say). Then $T$ decreases gradually and again approaches to zero
asymptotically. Thus a tiny black hole will have zero Hawking
temperature, i.e., vanishing surface gravity. As the size of the
event horizon increases, the Hawking temperature as well as the
surface gravity increases and reaches a maximum value at the event
horizon radius $r_{+2}$ Then further increase of the radius of the
event horizon results a decrease of Hawking temperature (and
surface gravity) and finally infinite black hole will have zero
temperature and vanishing surface gravity.

The figure $1(b)$ shows that $c_{\omega}$ has two distinct
branches. Starting from zero, $c_{\omega}$ remains positive for
$r_{+1}<r_{+}<r_{+2}$. At $r=r_{+2}$, $c_{\omega}$ blows up and
changes sign. For $r_{+}>r_{+2}$, $c_{\omega}$ remains negative.
From negative infinity value $c_{\omega}$ increases sharply
reaches a maximum and then gradually decreases.

The graph for $F$ plotted against $r_{+}$ (in figure $(1c)$) shows
that free energy is positive except for small $r_{+}(<r_{+1})$.
The curve has a point of inflexion initially concave downwards and
then concave upwards.

Also the $F-T$ variation in figure $(1d)$ shows that there are two
branches of the curve. For one of the branches the free energy
increases very slowly with the increase of the Hawking temperature
to the maximum limit $T_{M}$ while for the other branch $F$ starts
from infinite value decreases sharply till $T=T_{M}$. Here
$T=T_{M}$ is a cusp type double point.

Finally, we note that $r_{+1}=\frac{1}{\sqrt{2\omega}}$ is the
minimum possible radius of the event horizon. At $r_{+}=r_{+2}
$(the positive real root of
$1+5r_{+}^{2}\omega-2r_{+}^{4}\omega^{2}=0$) there is a possible
phase transition. Here $c_{\omega}$ blows up and changes sign
(from positive to negative), the Ruppeiner metric becomes
degenerate and in $F-T$ diagram it corresponds to a cusp type
double point. In analogy with second order Hawking-Page (1983)
type phase transition (which occurs in the AdS or asymptotically
AdS black hole) the black hole in HL gravity (which asymptotically
behaves like Minkowskian one) admits a transition from initial
stable phase to an unstable one.

Further, at $r_{+}=r_{+1}=\frac{1}{\sqrt{2\omega}}$, the vanishing
of the Hawking temperature indicates the equality of the free
energy with the mass of the black hole and we have
$F=M=r_{+}=\frac{1}{2 \omega}$.

\section{Black hole Thermodynamics in Horava-Lifshitz Gravity with Entropy-II }
This section analyze the thermodynamical parameters with
entropy-$II$. Although, the entropy area relation holds in GR but
still it is interesting to study the black hole thermodynamics
with entropy-$II$ in HL gravity theory for a comparative study as
HL gravity corresponds to (with some restrictions) to GR in IR cut
off. In this section, one may note that recently Wang et al (2009)
have studied the first law of thermodynamics in IR modified
HL-space time using entropy-$II$.

\subsection{Calculations}

From equation $(14)$ we have horizons at
\begin{equation}
\bullet~~~~~r_{\pm}=M\pm\sqrt{M^{2}-\frac{1}{2\omega}}
\end{equation}
The surface area of event horizon is given by
$$A=r_{+}^{2}\int^{\pi}_{\theta=0}\int^{2\pi}_{\phi=0}
\sin\theta~ d\theta~ d\phi =4\pi r_{+}^{2}$$ So entropy of the the
black holes from equation $(16)$ has the form
\begin{equation}
\bullet~~~~~S=\pi r_{+}^{2}
\end{equation}
and from equations $(28)$ and $(29)$ we have,
\begin{equation}
\bullet~~~~~M=\frac{1}{2\sqrt{\pi}}\left[S^{\frac{1}{2}}+\frac{\pi}{2\omega}S^{-\frac{1}{2}}\right]
\end{equation}
\begin{equation}
\bullet~~~~~T=\left(\frac{\partial M}{\partial S}\right)_{\omega}
=\frac{2\omega S-\pi}{8\sqrt{\pi}\omega S^{\frac{3}{2}}}
\end{equation}

\begin{figure}
\includegraphics[height=3in,
width=3in]{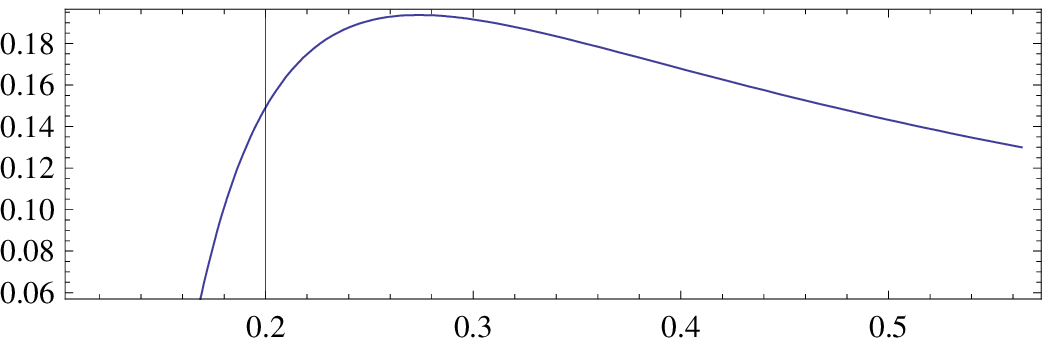}
\includegraphics[height=3in, width=3in]{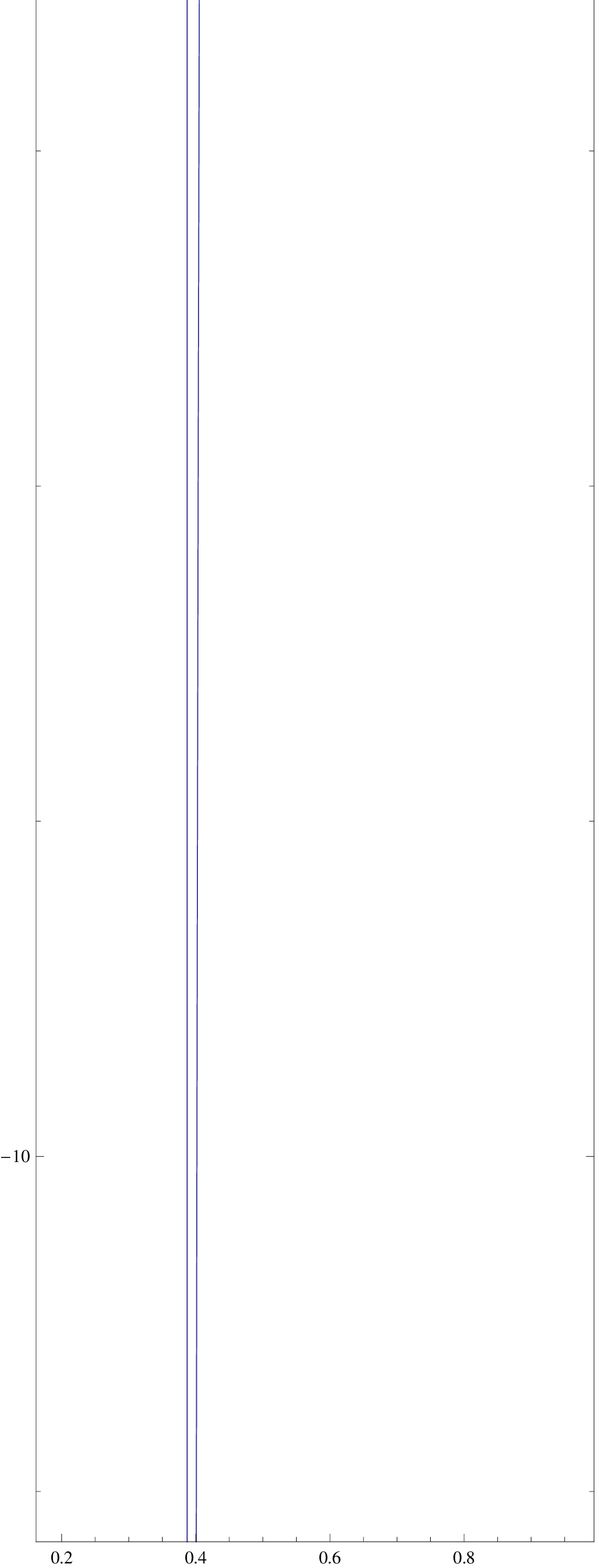}
Fig.(2a)~~$T$ vs $r_{+}$ graph for
$\omega=20$.~~~~~~~~~~~~~~~~Fig.(2b)$c_{\omega}$ vs $r_{+}$ graph
for $\omega=10$~~\\

\vspace{5mm}
\includegraphics[height=3in, width=3in]{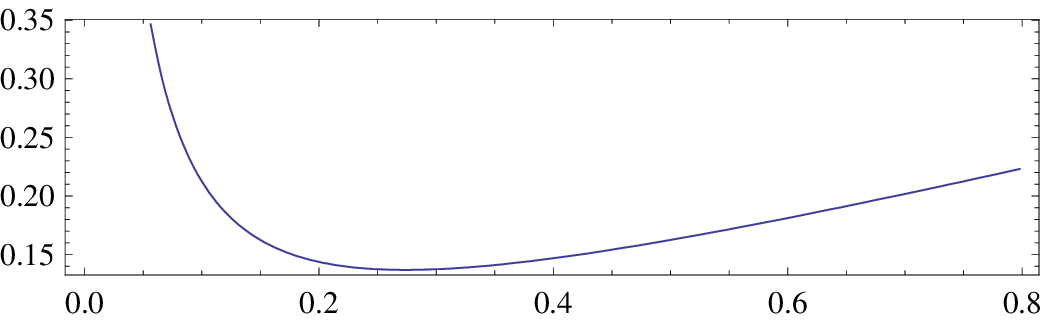}
\includegraphics[height=3in, width=3in]{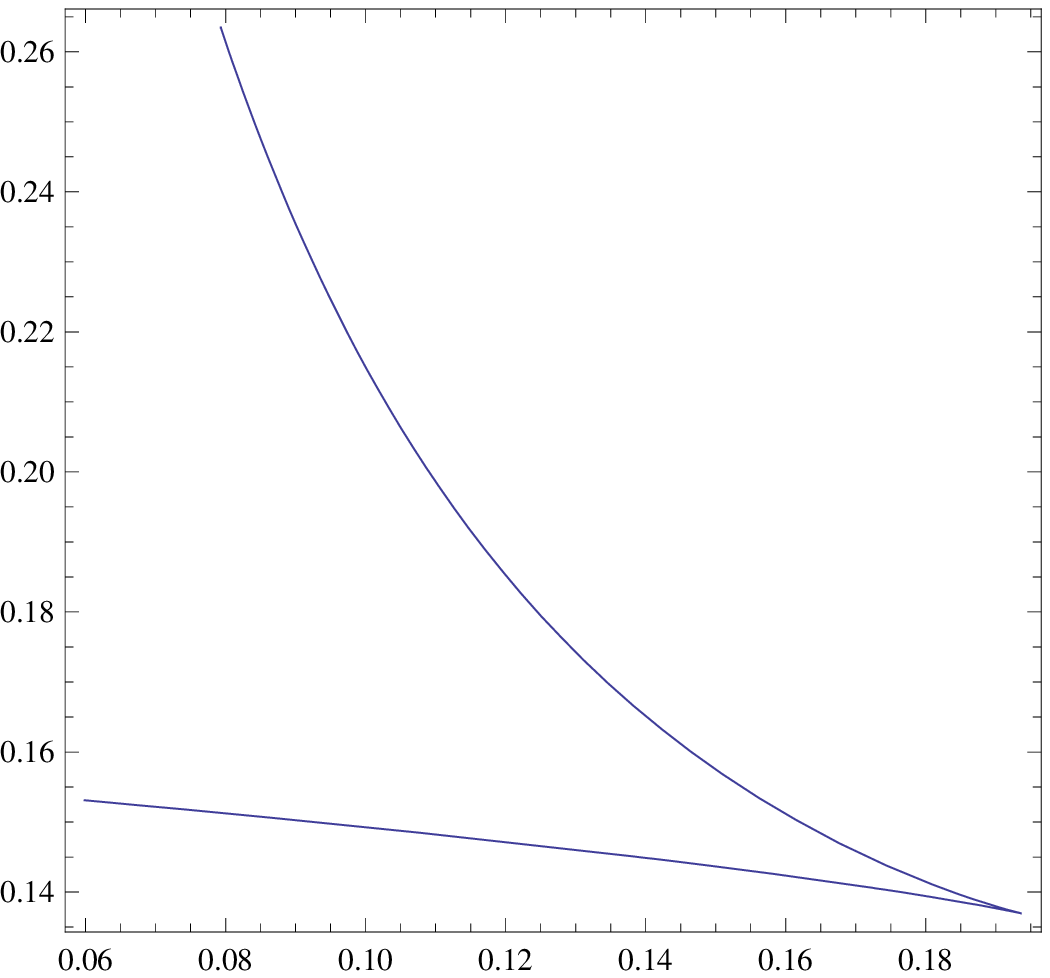}\\
Fig.(2c)~~$F$ vs $r_{+}$ graph for $\omega$
=20~~~~~~~~~~~~~~~~Fig.(2e)~~$F$ vs $T$ graph, $\omega$ =
20.~~~~~~~~~~~~~~~~~~~~~~~~\\

\hspace{1cm} \vspace{1mm} Fig. 2(a), 2(b) and 2(c) show the
variation of $T$, $c_{\omega}$ and $F$ respectively with $r_{+}$.
Fig 2(d)represents variation of $F$ with $T$ \vspace{5mm}
\end{figure}

\begin{equation}
\bullet~~~~~c_{\omega}=T\left(\frac{\partial S}{\partial
T}\right)_{\omega}=\frac{2S\left(2S\omega-\pi\right)}{\left(3\pi-2S\omega\right)}
\end{equation}
\begin{equation}
\bullet~~~~~Free ~~energy ~~(F)=M-T.S=\frac{1}{8\omega\sqrt{\pi}
S^{\frac{1}{2}}}\left(3\pi+2\omega S\right)
\end{equation}
In such case Weinhold metric is as follows
\begin{equation}
dS_{W}^{2}=\frac{\left(3\pi-2S\omega\right)}{16\sqrt{\pi}S^{\frac{5}{2}}\omega}dS^{2}+
\frac{\sqrt{\pi}dS~d\omega}{4S^{\frac{3}{2}}\omega^{2}}+
\frac{\sqrt{\pi}d\omega^{2}}{2S^{\frac{1}{2}}\omega^{3}}
\end{equation}
Ruppeiner metric has the form,
\begin{equation}
dS_{R}^{2}=\frac{2S\omega-3\pi}{2S\left(\pi-2\omega
S\right)}dS^{2}+ \frac{2\pi
dS~d\omega}{\omega\left(2S\omega-\pi\right)}+ \frac{4\pi
Sd\omega^{2}}{\omega^{2}\left(2S\omega-\pi\right)},
\end{equation}
which after diagonalization becomes
\begin{equation}
dS_{R}^{2}=\frac{\pi S-3\pi \omega^{2}+2 S\omega^{3}}{2S\left(\pi
\omega^{2}-\pi S-2S\omega^{3}\right)}dS^{2}+d\sigma^{2}.
\end{equation}
In this case $\sigma$can be determined by the equation
\begin{equation}
d\sigma=\frac{1}{2\omega}\frac{\pi^{\frac{1}{2}}}{S^{\frac{1}{2}}\left
(2S\omega-\pi\right)^{\frac{1}{2}}}
\left\{8S~d\omega~+\omega~dS\right\}
\end{equation}

\subsection{Graphical Analysis and Physical Interpretations}

The graphical representation of the thermodynamical quantities in
figures $(2a)$-$(2d)$ show very similar behavior as in previous
case. The figures for $T$ and $c_{\omega}$ are identical to the
earlier ones. Here also the minimum value of the radius of the
event horizon is $r_{+1}=\frac{1}{\sqrt{2\omega}}$ at which both
$T$ and $c_{\omega}$ are zero. $T$ has the maximum value
(=$T_{M}^{*}$) at $r_{+}=\sqrt{\frac{3}{2\omega}}$ where
$c_{\omega}$ diverges and changes sign.

The graph of $F$ against $r_{+}$ in figure $(2c)$ is distinct from
that in the last section. Although, in both cases $F$ is positive
throughout but previously $F$ is always
 an increasing function with a point of inflexion while in the present case
  $F$ starts from an extreme large value, decreases
sharply till a minima and then increases gradually.

The variation of $F$ against $T$ is shown in figure $(2d)$. As in
the previous case $F$ has two branches with a cusp at
$T=T_{M}^{*}~(i.e.,~r_{+}=\sqrt{\frac{3}{2\omega}})$. Hence in
this case there is also a phase transition at
$r_{+}=\sqrt{\frac{3}{2\omega}}$ of Hawking-Page type (with
degenerate Ruppeiner metric) and the black hole finally becomes an
unstable one.

\section{Conclusion}
We have studied the thermodynamics of the black holes in modified
Ho$\breve{r}$ava Lifshitz gravity (known as KS black hole) for two
choices of entropy - one classical and the other is a topological
one. For both choices, the black hole is initially stable and
becomes unstable through a Hawking page type phase transition. The
free energy $F$ is positive
 in both the cases supportingly the unstable nature of the black hole.
 However, the physical significance of the point of inflexion for the
 curve of $F$ in figure $1(c)$ is not clear. The maximum temperature representing the cusp
 type double point in the $F-T$ curve corresponds to phase transition -
  whether a general feature or not - is not known. Although, the thermodynamics of this KS- black hole has similarity with RN black hole but the thermodynamic geometries are quite distinct. The KS black hole has non-vanishing
   curvature indicating the statistical system to be interacting. Finally, we
  conclude that the black hole thermodynamics in modified Ho$\breve{r}$ava Lifshitz
  gravity for the above two choices of the entropy function is more or less identical.
  Lastly, the large black hole in modified Ho$\breve{r}$ava Lifshitz gravity is unstable globally while in
  Einstein gravity Schwartzschild AdS black hole is a stable one.
  \\\\

{\bf References :}\\\\
Arnowitt,  R. L. , Deser, S. ,Misner,  C. W. :{\it The dynamicsof general relativity, 'gravitation : an introduction to current research, L.Witten ed.'} (Wiley 1962), Chapter 7, pp227-265, arXiv:gr-qc/0405109.\\
Bardeen, J. M., Carter, B., Hawking, S. W. :{\it Commun. Math. Phys.} {\bf 31}, 161(1973).\\
Bekenstein, J. D. : {\it Phys. Rev. D} {\bf 7}, 2333(1973)\\
Cai, R. G.,  Cao, L. M., Otha, N. : {\it Phys. Rev. D} {\bf 80}  024003 (2009).\\
Cai, R. G., Cao, L. M.,Otha, N. : {\it Phys. Lett. B} {\bf 679} 504,(2009)\\
Calcagni, G. : {\it JHEP }  {\bf 0909} 112,(2009).\\
Catillo, A., Larra$\tilde{n}$aga,  A. : {\it arXiv:} {\bf 0906.4380v2} [gr-qc]\\
Davies, P. C. W. :{\it Proc. Roy. Soc. Lond.~A} {\bf 353}, 499(1977); 
Davies, P. C. W. :{\it Rep. Prog. Phys.} {\bf41},1313 (1977).\\
Davies, P. C. W. : {\it Class. Quant. Grav.} {\bf 6}, 1909(1989).\\
Hawking, S. W. : {\it Commun. Math. Phys.} {\bf 43}, 199(1975)\\
Hawking, S. W., Page, D. N. :{\it Commun. Math. Phys.} {\bf 87}, 577(1983)\\
Horava., P. : {\it Phys. Rev. D} {\bf 079} 084008,(2009.)\\
Ho$\breve{r}$ava, P. : {\it JHEP} {\bf 0903}, 020(2009).\\
Ho$\breve{r}$ava, P. : {\it Phys. Rev. Lett.}  {\bf 102} 161301,(2009).\\
Hut, P. :{\it Mon. Not. R. Astron Soc.} {\bf180}, 379(1977).\\
Kehagias, A., Sfetsos, K. : {\it Phys. Lett. B.} {\bf 678}, 123(2009).\\
Kiritsis, E., Kofinas, G. : {\it Nucl.Phys.B} {\bf 821} 467,(2009).\\
Kluson, J. : {\it JHEP}  {\bf 0907} 079,(2009).\\
Lu, H., Mei, J., Pope, C. N. : {\it Phys. Rev. Lett.} {\bf 103} 091301,(2009).\\
Ruppeiner, G. : {\it Phys Rev. A} {\bf 20}, 1608(1979).\\
Ruppeiner,  G. :{\it Rev. Mod. Phys.} {\bf67}, 605(1995);
Ruppeiner,  G. :{\it Rev. Mod. Phys.} {\bf68}, 313(E)(1996)\\
Takahasi, T., Soda, J. : {\it Phys. Rev. Lett.}  {\bf 102} 231301,(2009).\\
Wang, M., Jing, J. , Ding, C., Chen, S. {\it arXiv :}{\bf 0912.4832v1}[gr-qc] (24 Dec, 2009)\\
Weinhold, F.  :{\it J. Chem. Phys.} {\bf63}, 2479(1975).\\

\end{document}